\documentclass[aip, reprint]{revtex4-2}

\usepackage[utf8]{inputenc}
\usepackage[T1]{fontenc}
\renewcommand{\figurename}{Fig.}

\newcommand{\unit}[1]{\ \text{#1}}
\usepackage{chemformula}
\usepackage{upgreek}
\newcommand*{\PtC}{\ch{MeCpPt(Me)3}}
\newcommand*{\permalloy}{\ch{Ni80Fe20}}
\newcommand{\inlineabstract}[1]{\textbf{#1}}

\newcommand{\draftfig}{false}
\usepackage{hyperref}
\hypersetup{colorlinks=true, linkcolor=black, citecolor=black, urlcolor=blue}
\usepackage{color}

\usepackage[normalem]{ulem}
\usepackage{enumitem, amssymb}
\newlist{todolist}{itemize}{2}
\setlist[todolist]{label=$\square$}
\usepackage{pifont}
\begin{document}

\title{Domain wall automotion in three-dimensional magnetic helical interconnectors}



\author{L. Skoric}
\email[Author to whom correspondence should be addressed: ]{ls604@cam.ac.uk}
\affiliation{Department of Physics, Cavendish Laboratory, University of Cambridge, JJ Thomson Ave, Cambridge CB3 0HE, UK}

\author{C. Donnelly}
\affiliation{Department of Physics, Cavendish Laboratory, University of Cambridge, JJ Thomson Ave, Cambridge CB3 0HE, UK}
\affiliation{Max Planck Institute for Chemical Physics of Solids, 01187 Dresden, Germany}
\author{A. Hierro-Rodriguez}
\affiliation{SUPA, School of Physics and Astronomy, University of Glasgow, Glasgow G12 8QQ, UK}
\affiliation{Depto. Física, Universidad de Oviedo, 33007 Oviedo, Spain}
\author{M. A. Cascales Sandoval}
\affiliation{SUPA, School of Physics and Astronomy, University of Glasgow, Glasgow G12 8QQ, UK}

\author{S. Ruiz-Gómez}
\affiliation{ALBA Synchrotron, 08290 Cerdanyola del Vallès, Spain}
\author{M. Foerster}
\affiliation{ALBA Synchrotron, 08290 Cerdanyola del Vallès, Spain}
\author{M. A. Niño Orti}
\affiliation{ALBA Synchrotron, 08290 Cerdanyola del Vallès, Spain}
\author{R. Belkhou}
\affiliation{SOLEIL Synchrotron, L'ormes des Merisiers, Saint Aubin BP-48, 91192 Gif-Sur-Yvette Cedex, France}
\author{C. Abert}
\affiliation{Faculty of Physics, University of Vienna, Vienna, Austria}
\affiliation{Research Platform MMM Mathematics-Magnetism-Materials, University of Vienna, Austria}
\author{D. Suess}
\affiliation{Faculty of Physics, University of Vienna, Vienna, Austria}
\affiliation{Research Platform MMM Mathematics-Magnetism-Materials, University of Vienna, Austria}

\author{A. Fern\'andez-Pacheco}
\email{amaliofp@unizar.es}
\affiliation{Insituto de Nanociencia y Materiales de Aragón (INMA). CSIC-Universidad de Zaragoza, 50009 Zaragoza, Spain}

\date{\today}

\maketitle 


\inlineabstract{
    The fundamental limits currently faced by traditional computing devices necessitate the exploration of new ways to store, compute and transmit information \cite{waldropChipsAreMoore2016, zhangSystemHierarchyBraininspired2020}.
    Here, we propose a three-dimensional (3D) magnetic interconnector that exploits geometry-driven automotion of domain walls (DWs), for the transfer of magnetic information between functional magnetic planes.
    By combining state-of-the-art 3D nanoprinting\cite{fernandez-pachecoWriting3DNanomagnets2020} and standard physical vapor deposition, we prototype 3D helical DW conduits.
    We observe the automotion of DWs by imaging their magnetic state under different field sequences using X-ray microscopy\cite{kimlingPhotoemissionElectronMicroscopy2011, streubelEquilibriumMagneticStates2012,dacolObservationBlochpointDomain2014, jametQuantitativeAnalysisShadow2015, wartelleTransmissionXMCDPEEMImaging2017}, observing a robust unidirectional motion of DWs from the bottom to the top of the spirals. From experiments and micromagnetic simulations, we determine that the large thickness gradients present in the structure are the main mechanism for 3D DW automotion. We obtain direct evidence of how this tailorable magnetic energy gradient is imprinted in the devices, and how it competes with pinning effects due to local changes in the energy landscape. Our work also predicts how this effect could lead to high DW velocities, reaching the Walker limit \cite{schryerMotion180Domain1974} during automotion. This work  provides new possibilities for efficient transfer of magnetic information in three dimensions.
}

The exponentially increasing demands of the information age for denser, more efficient and better connected computing devices pose significant challenges to the microelectronics industry. Instead of relying purely on horizontal scaling, one way to address this is to start vertically stacking computing elements, a concept incorporated in the modern 3D V-NAND memories \cite{parkThreeDimensional128Gb2015}. In addition to offering higher densities, the move to 3D would offer new routes for higher integration and improved connectivity, enabling computing paradigms going beyond von Neumann architectures, such as neuromorphic computing \cite{markovicPhysicsNeuromorphicComputing2020}.

An area that would particularly benefit from the advance to 3D technologies is spintronics. While offering robust, low-power and non-volatile devices\cite{jiReliability8MbitEmbeddedSTTMRAM2019}, current 2D spintronic technologies are lacking in densities when compared to their CMOS counterparts\cite{dienyOpportunitiesChallengesSpintronics2019}.  On the other hand, the low power consumption of spintronic devices makes them particularly well suited to vertically integrated technologies where heat removal starts becoming problematic \cite{hanPowerThermalModeling2021}.
Going to 3D would allow us to leverage the many unique effects that arise in 3D nanomagnetic structures that could offer further scaling and increased functionality of computing elements \cite{fernandez-pachecoThreedimensionalNanomagnetism2017, streubelMagnetismCurvedGeometries2016}.
In order to achieve this, however, it is necessary to develop 3D interconnectivity in magnetic devices without requiring multiple energetically costly charge-to-spin conversion steps. It is thus important to find efficient mechanisms to transfer magnetic information between planes using entirely magnetic interconnectors.

\begin{figure}[t]
    \centering
    \includegraphics[draft=\draftfig, width=\linewidth]{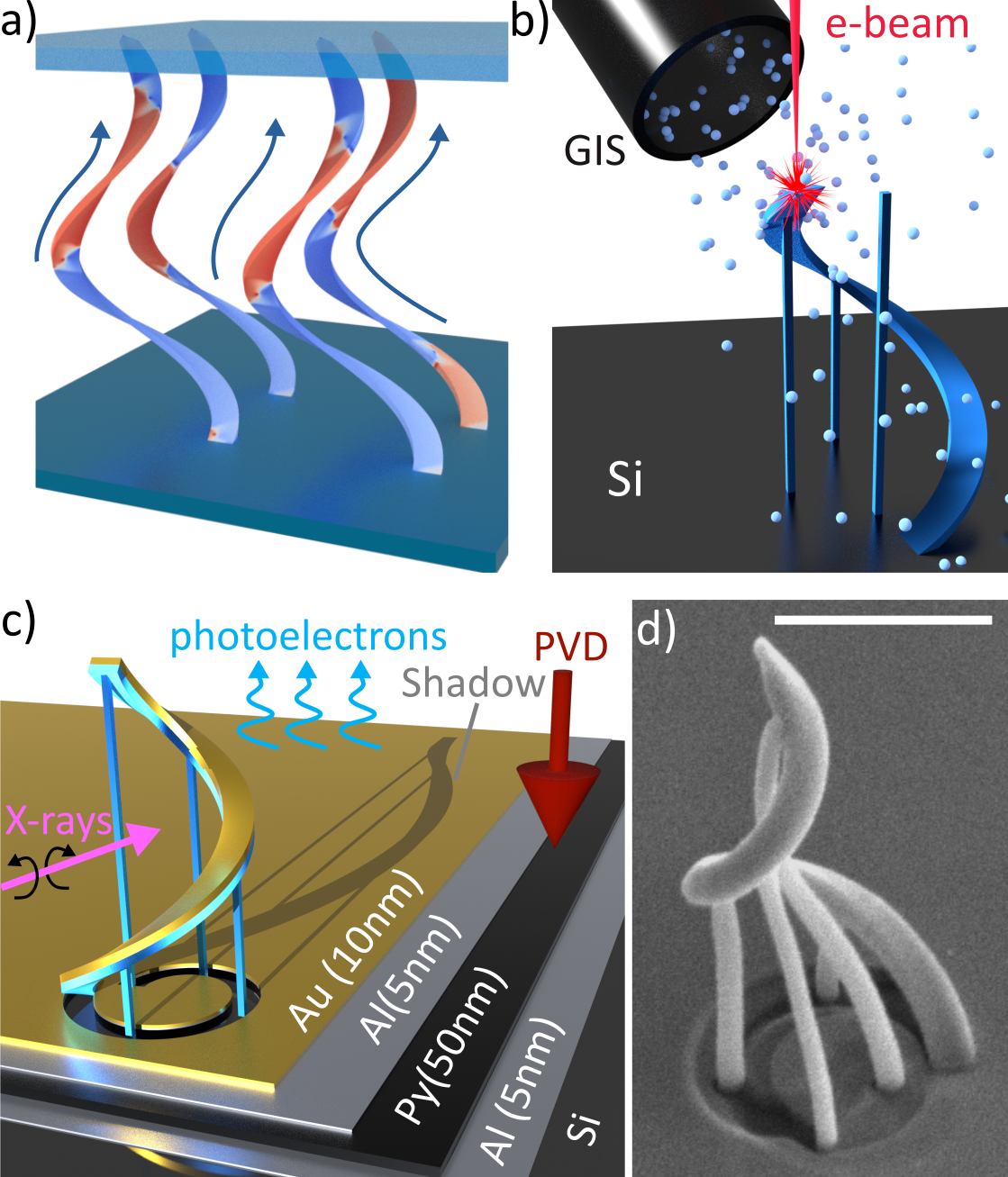}
    \caption{Fabrication of 3D magnetic interconnectors.
        (a) Concept device using 3D interconnectors to transfer magnetic information between two functional planes.
        (b) Scaffold for the DW conduit fabricated with FEBID using \PtC\ precursor on a Si substrate. The scaffold is a 3 $\upmu$m inwards-curving spiral with four supporting pillars for mechanical stability.
        (c) Functional films deposited using physical vapor deposition (PVD) onto the structure and the substrate. Due to the directionality of PVD, the supporting pillars and the area directly beneath the structure are free from PVD materials. The magnetic state of the structure is investigated with the  shadow X-ray magnetic circular dichroism photoemission electron microscopy (shadow-XPEEM) method by comparing the signal from right- and left-circularly polarized X-rays.
        (d) SEM image taken at 45° tilt of the resulting structure after X-ray irradiation is shown. A mild initial bending of the original FEBID scaffold when exposed to PEEM environment is observed, after which the structure becomes stable (see Supplementary Information S1 for details). The scale bar is 1\,$\upmu$m.
    }
    \label{fig:fabrication}
\end{figure}

Recently, it has been demonstrated that under the influence of external fields, DWs can be reliably injected from a plane, pinned, and propagated along 3D nanomagnetic conduits \cite{sanz-hernandezFabricationDetectionOperation2017}.
In addition to external stimuli, DWs can also be efficiently moved under the influence of geometry, driven by intrinsic spin-structure changes \cite{richterLocalDomainWallVelocity2016, nikonovAutomotionDomainWalls2014, mawassSwitchingDomainWallAutomotion2017}. This so-called DW automotion is a promising mechanism which could allow fast, low-power and robust transfer of magnetic information. In particular, using geometry-induced motion, the information could be transferred between functional planes where the state-of-the-art spintronic tools can be applied  (Fig. \ref{fig:fabrication}a).

While there are a number of ways in which the geometry of a system can be exploited to induce DW automotion, here we focus on three main effects: curvature gradients\cite{yershovGeometryinducedMotionMagnetic2018, yershovCurvatureinducedDomainWall2015}, changes in the cross-section of magnetic material \cite{mawassSwitchingDomainWallAutomotion2017, fernandez-roldanModelingMagneticfieldinducedDomain2019}, and magnetostatic interactions with magnetic surface charges.
Specifically, we design a prototype 3D magnetic nanostructure in order to investigate the viability of the geometry-driven DW motion using state-of-the-art 3D nanofabrication.

We realize our prototype 3D automotive devices by harnessing Focused Electron Beam Induced Deposition (FEBID) to deposit a smooth 3 $\upmu$m tall, 150 nm wide, spiral scaffold with non-magnetic C-Pt material (Fig. \ref{fig:fabrication}b) \cite{skoricLayerbyLayerGrowthComplexShaped2020}. The spiral has an increasing out-of-plane tilt, and smoothly curves inwards.
Following a previously developed fabrication procedure\cite{sanz-hernandezFabricationScaffoldBased3D2018,sanz-hernandezFabricationDetectionOperation2017}, we subsequently deposit functional materials with physical vapor deposition (PVD) perpendicular to the substrate onto the whole sample (Fig. \ref{fig:fabrication}c). For the magnetic layer, we use 50 nm of permalloy (\permalloy) due to its low coercive fields, and good DW conduit properties \cite{allwoodMagneticDomainWallLogic2005,nahrwoldStructuralMagneticTransport2010}. We sandwich the permalloy layer with 5 nm Al layers to prevent oxidation, and add a 10 nm Au capping layer that serves as a highly efficient source of photoelectrons in shadow-PEEM and suppresses the XMCD signal from the Py on the substrate \cite{wartelleTransmissionXMCDPEEMImaging2017}.

The fabricated conduit combines magnetostatic interactions, curvature and thickness gradients, all three of which are mechanisms for DW automotion as discussed earlier. Firstly, curvature is known to induce an effective Dzyaloshinskii-Moriya interaction (DMI) and anisotropy\cite{shekaNonlocalChiralSymmetry2020} that can lead to the DW automotion or pinning in curvilinear systems with inhomogeneous curvature \cite{yershovGeometryinducedMotionMagnetic2018, yershovCurvatureinducedDomainWall2015}. In our structure, this is implemented by the inwards curving spiral geometry, with a curvature gradient of 0.09 $\upmu$m$^{-2}$ in its central region (see Supplementary Information S2).
While the three-dimensional wires also contain torsion, it is expected to introduce negligible quadratic corrections to the automotion, unless the motion is also driven by spin torques \cite{yershovGeometryinducedMotionMagnetic2018,yershovCurvatureTorsionEffects2016}.
Secondly, changes in the cross-sectional area are known to strongly affect the DW energy landscape, preferentially moving it towards the thinner region \cite{fernandez-roldanModelingMagneticfieldinducedDomain2019}. Taking advantage of the directionality of evaporation, by the increasing steepness of the structure we induce a negative thickness gradient, and thus a decreasing cross-sectional area normal to the spiral surface with height. Based on the model of the fabricated spiral, an average thickness gradient of -5.3 nm/$\upmu$m is obtained in this way (see Supplementary Information S2).
In 2D films, thickness gradients are often achieved using moving shutters\cite{petrSurfacesRoughnessGradient2016, vishwakarmaStudyInterfacesHf2020} or plasma enhanced chemical vapor deposition methods\cite{lopez-santosMicronscaleWedgeThin2017} which create wedge thin films on a scale of tens of microns to millimeters. Thus, the highly localized thickness gradients achieved with 3D nanopatterning present  a powerful advantage of this three-dimensional fabrication procedure.

Finally, while the bottom of the spiral is connected to the substrate film, the top is freely standing which results in the formation of magnetic surface charges. These can interact with the DW charges, inducing its motion. The magnetostatic force between the charges scales with the distance of the DW from the edge $r$ as $1/r^2$. Therefore, these interactions are negligible in extended systems such as complex circuits where DWs are far from the edges. However, they are important to consider when investigating finite systems such as the ones in this work.

\begin{figure*}[t!]
    \centering
    \includegraphics[draft=\draftfig, width=0.85\linewidth]{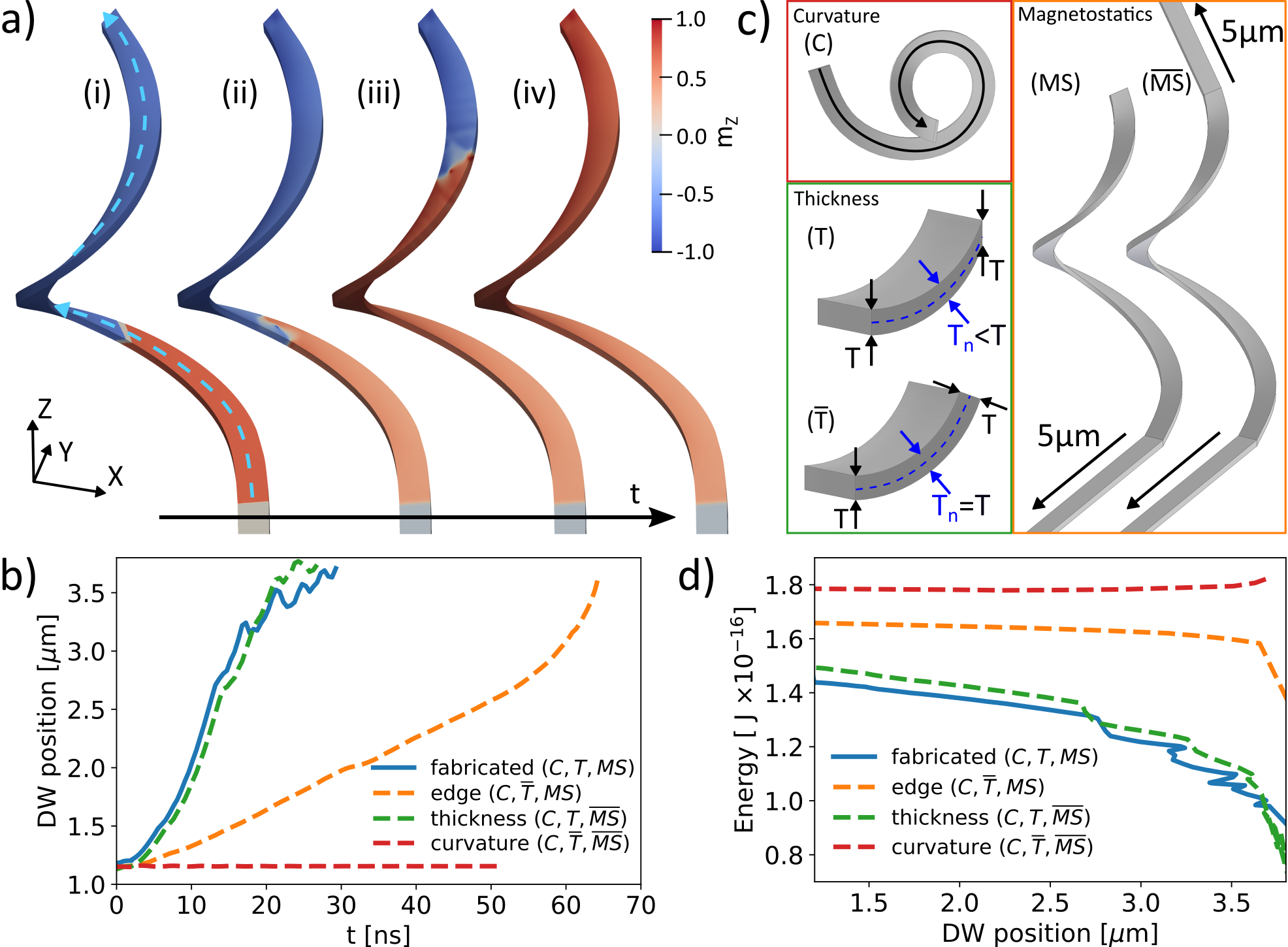}
    \caption{Simulations of DW automotion in 3D conduits.
        (a) Simulation snapshots for a model matching the fabricated structure: (i) initialization, (ii) relaxation into a DW, (iii) DW motion, (iv) final state. The blue dashed line on (i) shows the spiral central line along which the DW position is measured.
        (b) Position of the DW during simulations as a function of simulation time, with the as-fabricated structure shown in blue.
        (c) The relevance of different automotion effects is investigated by considering a spiral with curvature gradients ($C$), with ($T$) and without ($\overline{T}$) gradients in thickness normal to the spiral ($T=50\unit{nm}$), and with ($MS$) and without ($\overline{MS}$) surface charges at the top edge, the effect of the charges being removed by adding a 5 $\upmu$m extension. All structures have a 5 $\upmu$m extension at the bottom edge to simulate the continuous connection with the substrate.
        The curvature gradient (red line in (b)) is not enough to drive the DW motion by itself. While the edge magnetostatics (orange line in (b)) can drive the motion, the strongest effect is achieved by thickness gradients (green line in (b)). After 10 ns speeds of 200 m/s are surpassed, with the DW exhibiting Walker breakdown and the characteristic oscillatory motion.
        (d) Energy evolution of the DW as a function of position on the spiral during motion. The edge magnetostatics (orange) driven motion shows a slow, steady decrease in energy that becomes steeper as DW approaches the top. The thickness gradient driven motion has throughout a more steeply decreasing energy with step-like drops in the Walker regime.
        For the case with ``only curvature'' (red), where no automotion is observed, the DW is initialized at different positions on the spiral and its energy measured upon relaxation.
    }
    \label{fig:simulations}
\end{figure*}


To investigate the effect of the geometry on the DW automotion and determine the dominant automotive force in our system, we first perform dynamic finite-element micromagnetic simulations (see Methods). We use the model of the 3D structure closely matching the investigated spiral (see Supplementary Information S2). Starting from the magnetization in a head-to-head configuration (Fig. \ref{fig:simulations}a, part i), we observe the formation of a vortex DW (Fig. \ref{fig:simulations}a, part ii), which moves up purely under the influence of the geometry (Fig. \ref{fig:simulations}a, part iii), and ultimately fully switches the structure (Fig. \ref{fig:simulations}a, part iv). As the DW moves through the structure (blue line in Fig. \ref{fig:simulations}b), it accelerates towards the top, reaching speeds above 200 m/s (see Supplementary Information S3 and S4) before exhibiting the Walker breakdown. The characteristic Walker breakdown-induced oscillations \cite{schryerMotion180Domain1974} can be seen by the back-and-forth motion of the DW after $t=15\unit{ns}$ (see Supplementary Information S5 for details).

With such high velocities observed in the combined system, we consequently perform simulations on a modified model to determine the relative strengths of the driving mechanisms. Firstly, we focus only on the curvature gradient-driven automotion. We suppress the effect of thickness present in the real structure by modifying the film cross-section to obtain uniform thickness (thickness panel in Fig. \ref{fig:simulations}c). Moreover, we suppress any significant magnetostatic interactions with the edge by adding a 5 $\upmu$m extension (magnetostatics panel in Fig. \ref{fig:simulations}c). In the modified structure, the DW remains at the position where it was initialized, exhibiting no significant automotion (red line in Fig. \ref{fig:simulations}b). The weak effect of inhomogeneity of curvature is further reinforced by observing no significant changes in the energy of the system with DW initialized at different heights along the structure (red line in Fig. \ref{fig:simulations}d). This agrees with previous theoretical studies where an order of magnitude larger curvature gradients were used to induce automotion \cite{yershovGeometryinducedMotionMagnetic2018}.

We next introduce realistic edge magnetostatics by removing the top extension while keeping the model otherwise identical (magnetostatics panel in Fig. \ref{fig:simulations}c). This results in the DW moving steadily to the top, speeding up as it approaches the top of the structure (orange line in Fig. \ref{fig:simulations}b). However, the motion is slow ($\sim$ 30 m/s, see Supplementary Information S4) compared to the original structure, with only a weak gradient in energy as the DW moves up (orange line in Fig. \ref{fig:simulations}d).

Finally, we remove the edge magnetostatics by again adding the 5 $\upmu$m extension, and consider the influence of gradients in film thickness (thickness panel in Fig. \ref{fig:simulations}c).
In the simulated structure with thickness gradient (green line in Fig. \ref{fig:simulations}b), the DW accelerates to the top with motion closely matching the as-fabricated structure that combines all three automotive effects. Due to the lack of additional magnetostatic effects, the motion is slightly slower, and the Walker breakdown delayed.
Furthermore, the energy evolution of the purely thickness driven motion (green line in Fig. \ref{fig:simulations}d) approximately matches the fabricated structure (blue line in Fig. \ref{fig:simulations}d), with the Walker breakdown accompanied by the release of spin waves, resulting in the step-like drops in energy. From this study combining different simulations, we conclude that the DWs in our structures are predominantly driven by the spatial modulation of film thickness, while the curvature gradients and the edge magnetostatics are secondary effects.
It is worth noting that such a local gradient in thickness is difficult to realize in 2D devices.

\begin{figure*}[t!]
    \centering
    \includegraphics[draft=\draftfig, width=\linewidth]{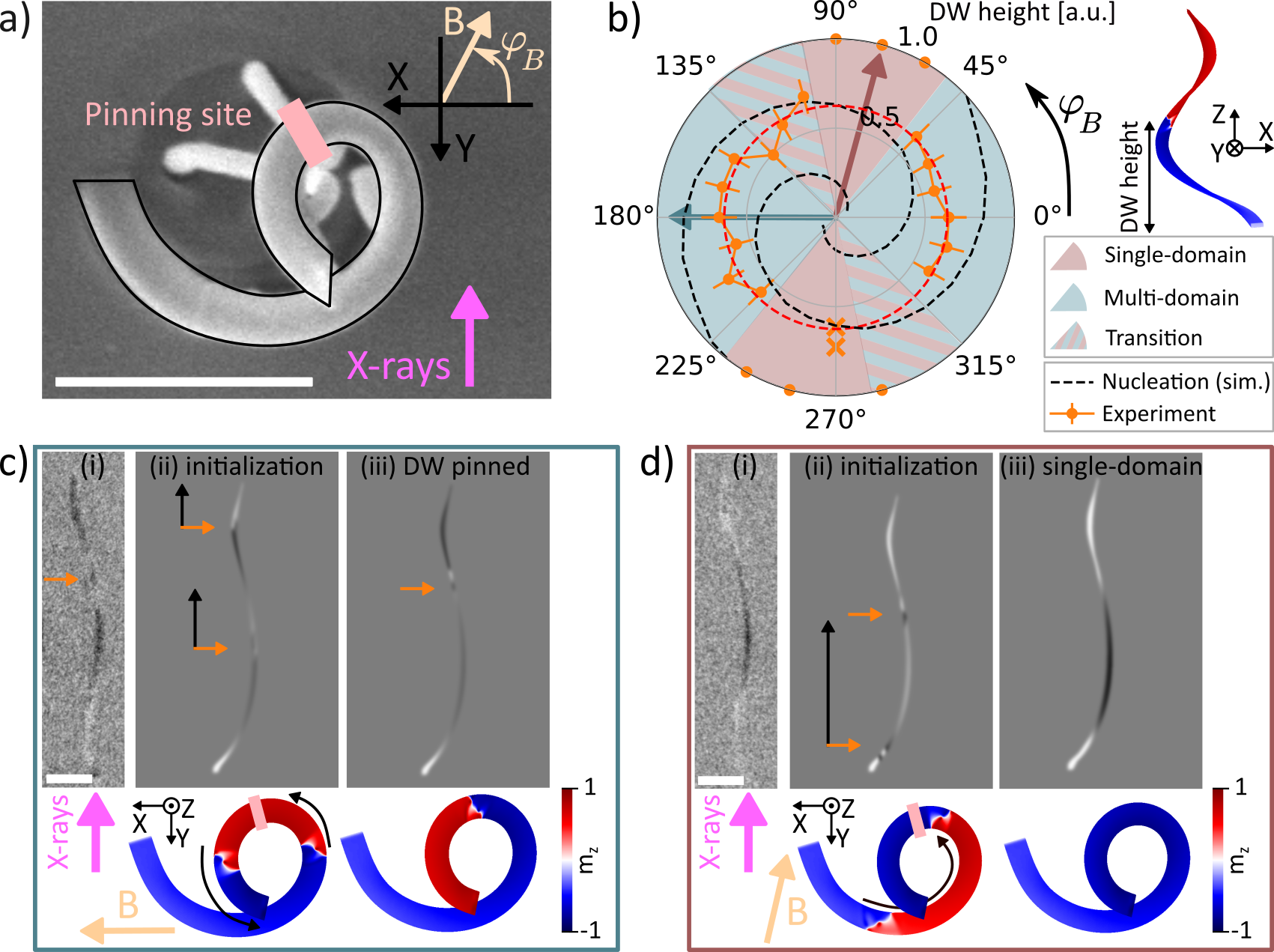}
    \caption{Angular study.
        (a) Top view of the fabricated structure with outline to guide the eye. The definition of the coordinate system is shown in top right, and the direction of X-rays denoted in pink. The location of the potential pinning site is shown in light pink.
        (b) Polar plot of the DW height normalized by the total structure height as a function of the initialization field angle $\varphi_B$  (see Supplementary Information S9 for measurement details). Orange symbols correspond to experimental data, after removal of the initializing field and the following automotion. The black dashed lines represent the nucleation site (before automotion) of DWs as predicted by simulations. In the pink shaded region we experimentally observe single-domain final states denoted by the points at $r=1$ (see (c) for example). In the blue-shaded region, we measure multi-domain final states with the DW position largely independent of the initialization angle (see (d) for example), suggesting a pinning site approximately 63\% up the structure (red dashed line). The striped region indicates the transition region where both types of behaviors can be observed. At the points denoted by crosses, we observe states with DWs that are not fully-annihilated (see Extended Data Fig. \ref{fig:angles_full}). The arrows mark the two initialization directions shown in more detail in (c) and (d).
        (c) Initialization at 180° with corresponding simulations. Experiment (i) shows the DW in the upper part of the structure (orange arrow). X-rays are coming from below (pink arrow). Simulations (ii, iii) show the top view of the structure state colored by $m_z$ (bottom), and the corresponding PEEM shadow (top). After the initialization at 180° (ii), both DWs (orange arrows) move up (black arrows) with the lower one stopping at the pinning site, and the upper one escaping through the top (iii).
        (d) Initialization at 60° leads to single-domain state (i). Here, both DWs (orange arrows) are initialized below the pinning site (ii), and annihilated to reach the single-domain state (iii). All scale bars are 1 $\upmu$m.
    }
    \label{fig:angles}
\end{figure*}



Having determined the dominant mechanism for DW automotion in the 3D interconnectors under investigation, we next use shadow X-ray magnetic circular dichroism photoemission electron microscopy (shadow-XPEEM) to experimentally investigate the automotion predicted by simulations. In shadow-XPEEM, the photoelectrons excited by the X-rays are measured in the shadow of the structure, exploiting X-ray magnetic circular dichroism (XMCD) to probe the magnetization parallel to the incident beam  \cite{kimlingPhotoemissionElectronMicroscopy2011, wartelleTransmissionXMCDPEEMImaging2017}. Since our structure has a complex 3D shape, interpreting the contrast is not trivial. We thus supplement the experimental results by computing the resulting XMCD images of the micromagnetic simulations (see Methods).

In order to observe DW automotion experimentally, we measure the magnetic states of the spirals and the location of DWs as a response to magnetic fields. Specifically, we perform two types of experiments. First, in an ``angular study'', we test the whole device functionality by generating DWs at different positions within the structure. For this, we apply an in-plane saturating magnetic fields at different angles, and measure the remanent state for each angle. Secondly, after understanding the behavior at multiple angles, we select a suitable angle for an ``initialize and release'' experiment, where we initialize a pair of DWs at a chosen angle (150°), and track their position as the magnetic field is reduced.


In the ``angular study'' we initialize the system with an in-plane saturating (70 mT) field in 15° increments, imaging the magnetic state of the spirals at remanence with a fixed X-ray direction (see Fig. \ref{fig:angles}a for the definition of the coordinate system). Depending on the field initialization angle, we observe two main regimes (Fig. \ref{fig:angles}b): single-domain between 45° and 105° (and by symmetry between 225° and 285°), and multi-domain between 315° and 45° (and by symmetry between 135° and 225°), with a narrow variable transition region between them. The multi-domain regime contains a single DW which is located at approximately the same place, between 50-70\% up the structure height for the whole range of angles (orange dots in Fig. \ref{fig:angles}b, see Supplementary Information S9 for measurement details). For each of the regimes, we show the PEEM images along one example angle: 180° for multi-domain (Fig. \ref{fig:angles}c), and 75° for single-domain (Fig. \ref{fig:angles}d).
This behavior was found to be reproducible across an array of three structures, with the full data in Extended Data Fig. \ref{fig:angles_full}.

To understand these experimental findings, we use simulations considering the state initialized by the applied magnetic field and the time evolution of the magnetization when the field is removed. Specifically, when the magnetic field is applied, two diametrically opposed DWs are formed (dashed lines in Fig. \ref{fig:angles}b and Extended Data Fig. \ref{fig:angles_full}), that in all cases show a smooth upwards automotion when the field is removed, ultimately leading to a single-domain state.

The experiments showing multi-domain states therefore reveal the presence of pinning sites due to imperfections in the structures that compete against DW automotion.
As the DWs initialized at different positions in the multi-domain regime are consistently measured at the same place following relaxation, this implies some have moved under no fields and have been pinned during automotion. Indeed, this final state is consistent with the systematic DW automotion up the structure after the field is removed, with a pinning site located approximately 63\% up the structure. This location corresponds to the connection to one of the supporting legs (see Fig. \ref{fig:angles}b), implying that the presence of this support introduced local changes in the magnetic energy landscape, most likely due to the changes in the strain of the film at that location.

\begin{figure*}[ht!]
    \centering
    \includegraphics[draft=\draftfig, width=\linewidth]{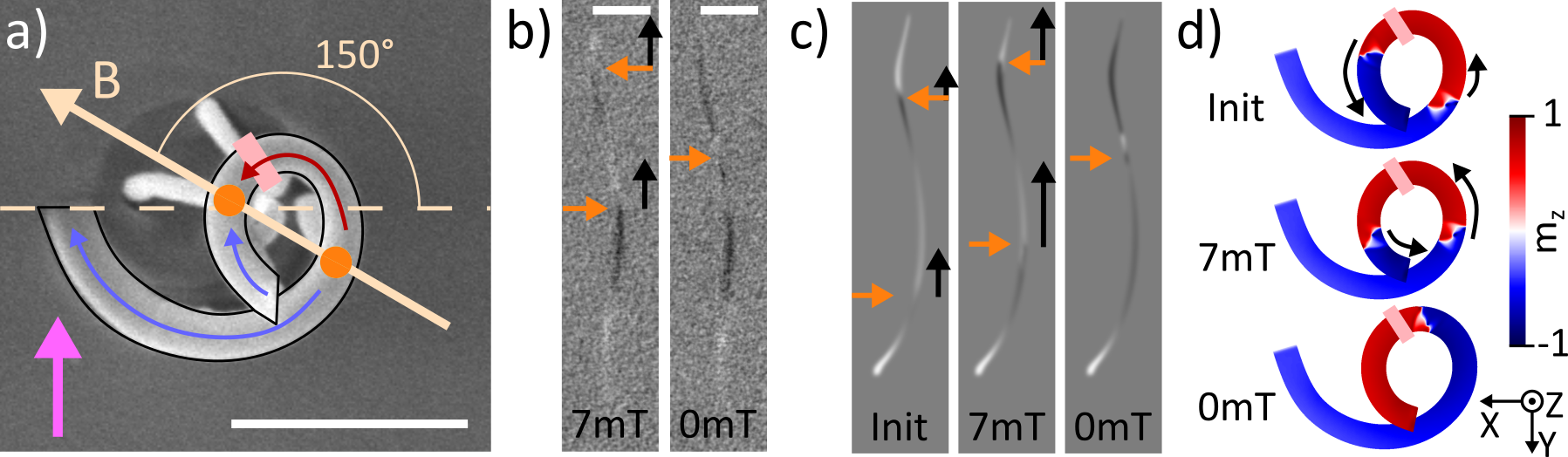}
    \caption{Initialize and release.
        (a) Top-view SEM image of the structure with outline to guide the eye. The direction of the applied field $B$ is shown in yellow, and the direction of X-rays in pink. The schematic of the expected magnetization state at high field values before automotion, with the location of domains (blue and red arrows), DWs (orange dots) and the pinning site (light pink) are overlaid onto the structure.
        (b) Shadow-XPEEM snapshots of the magnetization states at 7, 4, and 0 mT. All scale bars are 1 $\upmu$m.
        (c) Simulated shadows for: (0) as initialized DW, (1) relaxed state for 7 mT fields applied at 150° (corresponding to image 1 in (b)), (3) single DW simulation snapshot with the closest match to 0 mT (3) state in (b). Orange arrows in (b, c) indicate the positions of the DWs at each field value; black arrows indicate the upward automotion of the DWs observed between this and the following field value. (d) Top view of the simulation states colored by $m_z$ corresponding to the computed PEEM contrast in (c).}
    \label{fig:initrelease}
\end{figure*}

The experimental finding that some angles lead to multi-domain and others to single domain states can be understood by the location of the pinning site relative to the two initialized DWs.
For angles corresponding to the multi-domain regime, the two initialized DWs are on opposite sides of this pinning site (see how the two dashed lines in Fig. \ref{fig:angles}b are positioned in between the orange line). The DW initialized above the pinning site is expected to escape through the top, while the one below gets stuck below the barrier.
The same interpretation is also consistent with the single-domain regime. There, both DWs are initialized below the pinning site (see how both dashed lines in Fig. \ref{fig:angles}b are positioned below the orange line). As they move up, they are forced into each other by automotion, annihilating at the barrier (see simulations in Fig. \ref{fig:angles}d).
In particular, the upper DW gets trapped in the pinning site first, leading to the interaction of both walls when the second also reaches this area. In most cases, both walls annihilate, leading to a single domain state as observed in experiments (Fig. \ref{fig:angles}d).
However, in a few cases we observe how the two domain walls are still present at remanence, being located around the pinning site region (see crosses in Fig. \ref{fig:angles}b and associated images in Extended Data Fig. \ref{fig:angles_full}). The absence of annihilation is expected when the pinning of the lower wall is greater than its automotive effect, and also due to the possible presence of topological repulsion\cite{mawassSwitchingDomainWallAutomotion2017, lewisMeasuringDomainWall2009}.
Furthermore, variations of DW structure for speeds above the Walker breakdown, and the stochastic nature of DW pinning processes will also result in a transition region between single- and multi-domain states that is not sharp, with variances across the three measured structures (see Extended Data Fig. \ref{fig:angles_full}).


Following the angular study that allowed us to indirectly observe the effect of DW automotion and understand how DWs behave as a function of the direction of the initialization field, we next design an ``initialize and release'' experiment to directly track the automotion of the two DWs. For this, we initialize the structure with 70 mT fields at 150° (corresponding to the multi-domain regime, Fig. \ref{fig:initrelease}a), and image the magnetic configuration of the spiral as the field is reduced to zero in discrete steps, measuring at 7 mT, 4 mT and 0 mT (see Fig. \ref{fig:initrelease}b for the cases corresponding to the two extreme values). Complementary simulations following the same protocol are also performed (Fig. \ref{fig:initrelease}c,d).

At 7 mT, we observe a multi-domain state with the two DWs located below and above the pinning site, as in the initialization experiment discussed before (Fig. \ref{fig:initrelease}b). Both domain walls are observed above the positions where they were initialized according to simulations (Fig. \ref{fig:initrelease}c,d), indicating that the reduction of the field from 70 to 7mT has already allowed the automotion of DWs up the spiral. Indeed, when relaxing this simulated state under the application of 7 mT field, both DWs move up and against the field, in agreement with the experimentally observed state.

Reducing further the field from 7 mT to 4 mT does not introduce a noticeable change in the experimental state (not shown here), implying the automotion effect is smaller than the combined effect of pinning and the opposing external magnetic field. Removing the fields completely, in both the experiment and simulations, leads to the DWs moving up the spiral. We observe the upper DW escaping through the top, with the bottom one remaining pinned during automotion at approximately 55\% up the spiral, in good agreement with the position of the pinning site identified in the angular study.
The direct observation of systematic upward motion of the DWs further confirms that DWs in this type of 3D magnetic interconnectors experience an automotive force which is purely geometrical in character, resulting in a spontaneous unidirectional motion into the third dimension, and representing a new opportunity for low energy magnetic interconnects.


In summary, in this work, we demonstrate the realization of 3D complex-shaped DW interconnectors that exhibit geometrically driven domain wall automotion. These devices are fabricated with a combination of FEBID 3D nanoprinting with high-quality magnetic materials deposited with PVD. Via micromagnetic simulations, we evaluate the strength of individual contributions to the automotion and identify the gradients in film thickness as the dominant contribution, capable of moving DWs at speeds above 200 m/s and inducing Walker breakdown. We demonstrate the validity of the concept using shadow-XPEEM by directly observing the reproducible automotion and investigating the pinning landscape in the fabricated devices.
The realization of strong and localized gradients in thickness, which are difficult to realize in 2D, represent an attractive prospect for domain wall manipulation in three-dimensional architectures, and an interesting tool for functionalizing spintronic interconnectors.
The proposed concept could be exploited by scalable fabrication methods\cite{harinarayanaTwophotonLithographyThreedimensional2021}, opening the door to a field- and current-free method for unidirectional magnetic information transfer between functional planes.
This concept could help address two of the main challenges that spintronics faces when moving to three dimensions. Firstly, although the motion of domain walls is well controlled in 2D, robust motion in 3D interconnects has yet to be established. Secondly, current driven domain wall motion in 3D suspended nanostructures without large heat sinks faces severe heat dissipation challenges. By providing a purely geometric transfer of information between planes — an "elevator effect" — DW automotive devices such as the one described here promise to circumvent these challenges and offer a route to the implementation of 3D spintronic devices, that have possible applications in high-density magnetic memories and unconventional computing.


\bibliography{peem_spirals}
\clearpage

\renewcommand{\thefigure}{\thesection \arabic{figure}}
\renewcommand{\figurename}{Extended Data Fig.}
\setcounter{figure}{0}

\section*{Methods}

\subsection*{Fabrication}

The 3D spiral scaffold was fabricated with \PtC\ on p-doped Si substrate using Focused Electron Beam Induced Deposition with Helios 600 system at the Wolfson Electron Microscopy Suite of University of Cambridge. Electron beam was set to 21 pA, 30 kV. Beam scanning patterns were created from the designed STL files using the custom pattern generating software\cite{skoricLayerbyLayerGrowthComplexShaped2020}, and the total fabrication time was 14 minutes per structure.
Physical vapor deposition was done using an in-house thermal evaporator. The deposition rates were: 3.75 nm/min for Al, 1.8 nm/min for \permalloy, and 0.7 nm/min for Au. The structure imaged from multiple directions and further information on PEEM-induced deformation is available in Supplementary Information S1.

\subsection*{Micromagnetic simulations}

All micromagnetic simulations, were performed with finite-element magnum.fe library \cite{abertMagnumFeMicromagnetic2013}. We use the material parameters typical of permalloy: $M_S = 8\times 10^5 \unit{Am}^{-1}$, $A = 1.3 \times 10^{-11} \unit{Jm}^{-1}$ and consider the exchange field, the demagnetization field as well as the external field but no magneto-crystalline anisotropy.
The models were meshed with GMSH \cite{geuzaineGmsh3DFinite2009} with the characteristic mesh edge length of 5.7 nm. This corresponds to the permalloy dipolar exchange length $\ell = \sqrt{2A/\mu_0M_S^2} = 5.7 \unit{nm}$. The details of the simulation structure are available in the Supporting Information S2.

In order to acquire the simulation results presented in Fig. \ref{fig:simulations}, a single DW is initialized 1 $\upmu$m from the bottom in a head-to-head configuration with magnetization tangential to the spiral. Firstly, the state is relaxed into a DW by integrating Landau-Lifshitz-Gilbert equation (LLG) with high damping $\alpha=1$, simulating slow ramp-down of the fields after initialization. Secondly, the dynamics of the DW are investigated by again relaxing with LLG, now using realistic damping for permalloy $\alpha = 0.01$.

The simulations presented in the angles study (Fig. \ref{fig:angles}b, initializations in Fig. \ref{fig:angles}c,d, and in the Extended Data Fig. \ref{fig:angles_full}) were acquired by starting from the magnetization fully pointing in the direction of the corresponding magnetic field. The state is then integrated with LLG using high $\alpha=1$ and stopping once the DW is formed.

The simulations presented in Fig. \ref{fig:initrelease}c are acquired by starting from the magnetization pointing in the 150° direction, with a small (10°) tilt introduced to prefer left-handed VW circulation that better matches the data (see Supplementary Information S10 for details). As before, the state is first relaxed into a DW with high damping $\alpha=1$. The state is further integrated under 7 mT external fields with LLG using realistic $\alpha=0.01$ to allow DW dynamics (Fig. \ref{fig:initrelease}c, part 1). Finally, the effect of removal of the field is observed by removing the external field term, and further evolving the LLG (Fig. \ref{fig:initrelease}c, part 2).

\subsection*{Shadow-XPEEM}

Shadow-XPEEM images were taken at the CIRCE beamline of the Alba synchrotron facility \cite{aballeALBASpectroscopicLEEMPEEM2015}. The X-ray beam incident at 16° was set to $\sim 1\unit{eV}$ below Fe L$_3$ edge (see Supplementary Information S6 for XAS spectra). Start voltage was 6.0 V, and contrast aperture 30 $\upmu$m. In situ fields were applied using an improved version of the sample holder with quadrupole in-plane electromagnet described in Ref. \onlinecite{foersterCustomSampleEnvironments2016} which allows larger field values due to a reduced gap size in the magnetic yoke. For the details of image processing procedure, see Supplementary Information S7 and the corresponding software in Ref. \onlinecite{skoricXMCDpy2021}. The simulated images were generated from micromagnetic simulations using a custom ray tracing code in Ref. \onlinecite{skoricXMCDProjection2021}.


\subsection*{Measuring domain wall position}

In order to determine the variation of the DW position as a function of initial angle (Fig. \ref{fig:angles}b), the position of the DW in the shadow is measured from the bottom of the structure, normalized by the total shadow length. The location of the DW wall is acquired by plotting the XMCD signal as a function of position in the shadow and looking for regions of rapidly changing contrast, signifying the presence of the DW. For details, see Supplementary Information S10.

\begin{figure*}[t]
    \centering
    \includegraphics[draft=\draftfig, width=0.95\linewidth]{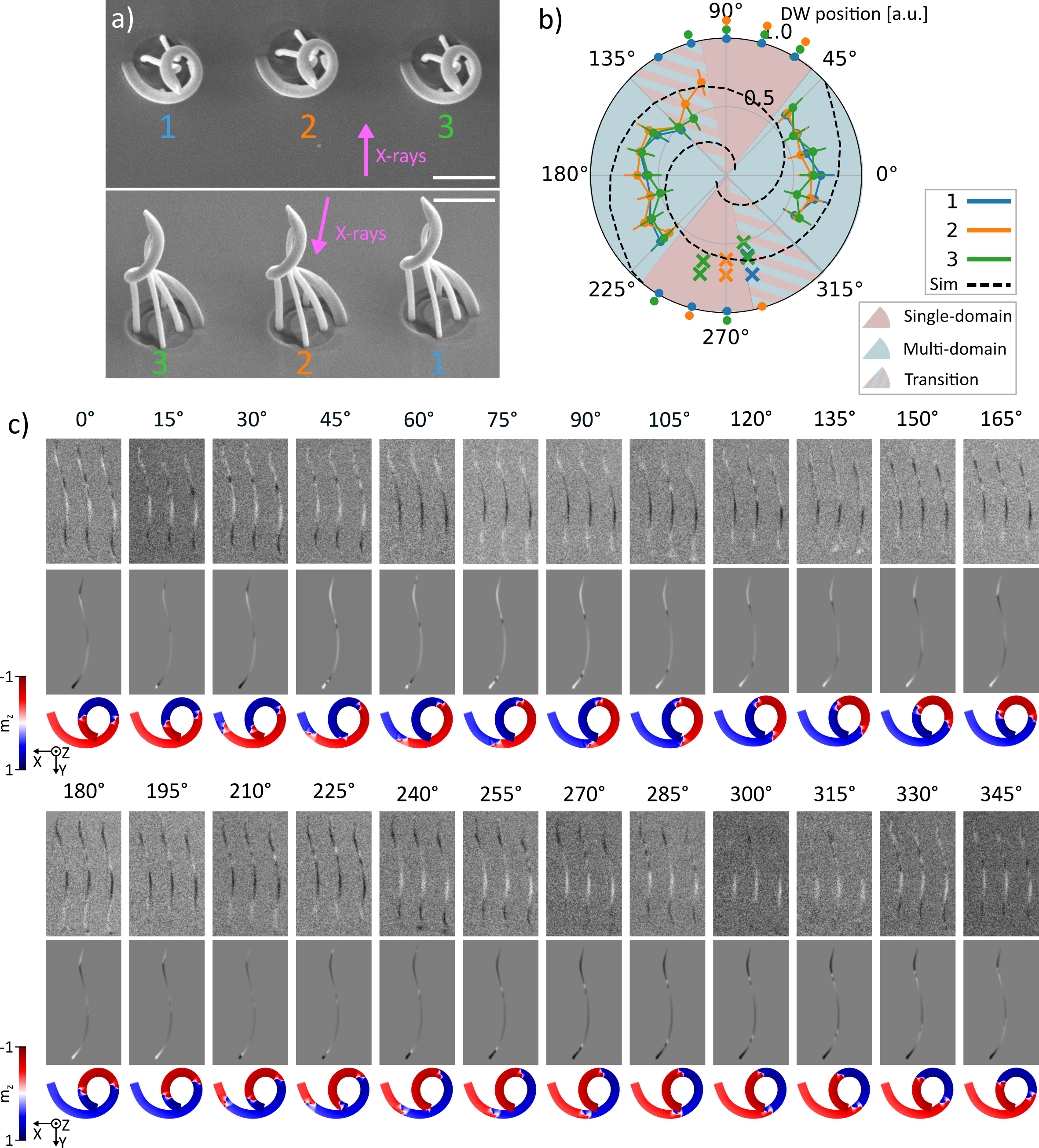}
    \caption{Angles study. (a) SEM images of the array of the three measured spirals: top view (top), and 45° tilt view (bottom). Scale bars are 1 $\upmu$m. The direction of X-rays is shown in pink.
        (b) Polar plot of the DW position extending the plot in Fig. \ref{fig:angles}b in the main text. The measured data for all three spirals is shown. In the main text, only results from the spiral 2 are shown.
        (c) The shadow-PEEM images of the system are shown after initialization at different angles with 15° increments. Below each image are the shadow of the simulated state immediately after the initialization (before automotion), and the top view of the simulated state colored by the $z$ component of magnetization.
        All observed states can be acquired by translating the simulated DWs upwards through automotion (see main text for details). The only exception are the states at 30° and 45°  and their inverses at 210° and 225° where in simulations, a formation of a small third domain at the bottom of the spiral is observed. This is likely due to a small disagreement between the fabricated model and the simulated structure near the connection with the substrate.
    }
    \label{fig:angles_full}
\end{figure*}

\section*{Supplementary Material}
See supplementary material for the technical information about the details of the fabricated structure, simulation model, and simulation and measurement data processing steps.

\begin{acknowledgments}
    The authors thank G. Divitini, and other staff of the Wolfson Electron Microscopy Suite for their technical support. We also appreciate insightful discussions with D. D. Sheka, R. P. Cowburn, P. Fischer and R. Streubel. AFP thanks the Universities of Cambridge and Glasgow, where part of this research was conducted.
    All data was measured at CIRCE (BL-24) beamline at ALBA synchrotron in collaboration with ALBA staff, proposal no. 2020024091. Preliminary measurements were performed at HERMES beamline at SOLEIL synchrotron in collaboration with SOLEIL staff, proposal nos. 20200879, 20190565 and 20170904.
    This work was supported by the EPSRC Cambridge NanoDTC EP/L015978/1, the Winton Program for the Physics of Sustainability, the project CALIPSOplus under Grant Agreement 730872 from the EU Framework Programme for Research and Innovation HORIZON 2020, and by the European Community under the Horizon 2020 Program, Contract no. 101001290, 3DNANOMAG.
    L. Skoric acknowledges support from St Johns College of the University of Cambridge.
    C. Donnelly  was supported by the Leverhulme Trust (ECF-2018-016), the Isaac Newton Trust (18-08) and the L'Or\'eal-UNESCO UK and Ireland Fellowship For Women In Science.
    A. Hierro-Rodriguez acknowledges support from Spanish AEI under project reference PID2019–104604RB/AEI/10.13039/501100011033.
    The authors acknowledge the University of Vienna research platform MMM Mathematics - Magnetism - Materials, and the FWF project I 4917.
\end{acknowledgments}

\section*{Data availability}
All data associated with this manuscript will be made openly available in a public repository when the manuscript is published.

\section*{Code availability}
The code used to produce XMCD images is available at \url{https://doi.org/10.5281/zenodo.5093811}, and the code for simulations of XMCD images from micromagnetic simulations at \url{https://doi.org/10.5281/zenodo.5094530}


\end{document}